\documentclass[reprint,aps,prc,twocolumn,superscriptaddress,floatfix,amsmath,amssymb,10pt]{revtex4-2}
\DeclareMathAlphabet{\mathpzc}{OT1}{pzc}{m}{it}
\usepackage{bm}
\usepackage{dcolumn}
\usepackage{amsthm}
\usepackage{amsmath}
\usepackage{amssymb}
\usepackage{graphicx}
\usepackage{braket}
\usepackage{xcolor}
\usepackage{fix-cm}
\usepackage{mathptmx} 
\usepackage[T1]{fontenc}
\usepackage[colorlinks,allcolors=blue]{hyperref}
\makeatletter
\def\NAT@def@citea{\def\@citea{\NAT@separator}}
\makeatother

\begin{document}

\title{Role of isospin composition in low energy nuclear fusion}

\author{Richard Gumbel}\email{richard.gumbel@vanderbilt.edu}
\altaffiliation{\textit{Current address:} Facility for Rare Isotope Beams, Michigan State University, East Lansing, Michigan 48824, USA}
\affiliation{Department of Life and Physical Sciences, Fisk University, Nashville, TN, 48823, USA}

\author{Christian Ross}\email{christian.ross@vanderbilt.edu}
\author{A.S. Umar}\email{umar@compsci.cas.vanderbilt.edu}

\affiliation{Department of Physics and Astronomy, Vanderbilt University, Nashville, TN, 37235, USA}

\date{\today}

\begin{abstract}
We employ a microscopic approach that examines the impact of isospin dynamics on the process of low energy nuclear fusion along an isotope chain and dependence on deformation. Our method utilizes the density constrained time-dependent
Hartree-Fock theory (DC-TDHF), where isoscalar and isovector characteristics of the energy density functional
(EDF) are examined in turn. This approach is applied to a series of fusion interactions of $^{176}$Yb with
increasingly neutron rich isotopes of Calcium. By evaluating the contributions from the isoscalar and
isovector components of the EDF, we look to quantify the influence of isospin composition on the conditions
under which fusion is most likely to take place. Our findings reveal that, in non-symmetric systems, the
isovector dynamics play a significant role.  It's typical effect is a reduction in the potential barrier,
which turns into enhancement for neutron-rich systems.
\end{abstract}

\maketitle

The study of fusion reactions is one of the major research areas
of low-energy heavy-ion physics~\cite{back2014,montagnoli2017,montagnoli2023}. Unfortunately, from the
theoretical standpoint the lack of a practical many-body approach
for sub-barrier tunneling requires the reduction of the fusion
studies to the determination of an effective ion-ion interaction
potential that allows for traditional tunneling methods to be
employed. If the ion-ion potential is initially computed with
frozen nuclear densities other quantal effects, such as the
excitation of the target and projectile, transfer of
nucleons during the initial phase of the collision
have to be included via various approximations.
The most commonly used method to achieve these goals is the
coupled-channels (CC) approach~\cite{hagino1999,hagino2012,hagino2022}. An alternate approach,
in which the dynamics of the collision is included at
the mean-field level is provided by the density-constrained
time-dependent Hartree-Fock (DC-TDHF) method~\cite{umar2006b,simenel2018}.

The dependence of fusion cross sections on neutron excess, or specifically the total isospin quantum number $T_z = (Z - N)/2$, is a significant question in the realm of fusion reactions, particularly fusion reactions involving exotic neutron-rich nuclei. This topic has gained further relevance, as rare isotope facilities conduct increasingly sophisticated exotic beam experiments~\cite{balantekin2014}. Furthermore, understanding the impact of isospin dynamics on fusion is crucial for the synthesis of superheavy elements using neutron-rich nuclei~\cite{loveland2007}.
Beyond its implications for nuclear structure and reactions, addressing this inquiry holds substantial importance for our comprehension of the nuclear equation of state (EOS) and symmetry energy~\cite{horowitz2014,li2014}, which is intimately related to nuclear structure~\cite{chen2015} and dynamics~\cite{danielewicz2002,tsang2009}, as well as most astrophysical phenomena~\cite{chamel2008,shen2011}.
Typically, the influence of isospin flow during heavy-ion reactions is discussed in terms of the $(N/Z)$ asymmetry of the target and projectile or the $Q$-values associated with nucleon transfer~\cite{jiang2014a}. However, there are still unresolved issues with the $Q$-value based
transfer methods, first the precise magnitude of fusion enhancement based on a known $Q$-value is not
well understood~\cite{jiang2015,liang2016}, second, for exotic nuclei $Q$-values may not be available, and finally the $Q$-value transfer
is based on the entrance channel properties of the participating nuclei whereas the dynamics during the neck formation phase of the
collision may introduce other dynamical effects. One such effect, the Pauli exclusion principle, has been recently
discussed~\cite{simenel2017,umar2021}. For reactions involving deformed nuclei the ion-ion barrier and the fusion dynamics
also depend on the orientation of the nuclei with respect to the beam axis~\cite{umar2006d,hagino2018,hagino2019}.

The time-dependent Hartree-Fock (TDHF) method supplemented with a density constraint, DC-TDHF, takes advantage of the dynamics
included in the TDHF time-evolution, which has been successfully utilized to study multinucleon transfer reactions~\cite{umar2008a,simenel2010,simenel2012b,sekizawa2019,wu2022},
deep-inelastic damped collisions~\cite{koonin1977,simenel2011,umar2017}, and quasifission~\cite{wakhle2014,umar2015a,umar2016,sekizawa2016,guo2018d,godbey2019,godbey2020,stevenson2022}.
The benefit of this approach is that both the structure
and reactions are handled on the same footing through an energy density functional with pre-determined parameters.
Hence, the dynamical transfer mechanism, and their influence on the ion-ion interaction potentials at the mean-field level
can be studied without making \textit{a priori} assumptions.

Within the TDHF theory, the totally antisymmetric many-body wavefunction is assumed to be a single
Slater determinant. Neglecting the two-body correlations preserves the Slater determinant
nature of the many-body state throughout the time evolution. This many-body state is
then used to construct the time-dependent action using an effective nucleon-nucleon
interaction. Variation of this action with respect to single-particle states
$\phi_{\lambda}^{*}$
\begin{equation}
\label{eq:variation}
    \frac{\delta S}{\delta \phi^{*}_{\lambda}} = \frac{\delta }{\delta \phi^{*}_{\lambda}} \int dt \braket{\Phi (t) | H - i\hbar \frac{\partial}{\partial t} |\Phi (t)} = 0,
\end{equation} gives us the most probable reaction path as a set of fully microscopic, coupled, nonlinear, self-consistent, time-dependent Hartree-Fock equations of motion for the single-particle states,
\begin{equation}
h(\{\phi_{\mu}\}) \ \phi_{\lambda} (r,t) = i \hbar \frac{\partial}{\partial t} \phi_{\lambda} (r,t)
\ \ \ \ (\lambda = 1,...,A)~,
\label{eq:TDHF}
\end{equation}
where $h$ is the single-particle Hamiltonian. Employing an effective interaction such as the
Skyrme interaction results in the total energy of the system to be represented as an
volume integral of an energy density functional~\cite{engel1975}
\begin{equation}
\label{eq:energy}
E = \int d^3\mathbf{r} {\cal{H}}(\mathbf{r})~.
\end{equation}
For the purposes of this work the Skyrme EDF may be decomposed into isoscalar and isovector parts~\cite{dobaczewski1995}
(in addition to the conventional kinetic and Coulomb terms) as:
\begin{equation}
\label{eq:edensity}
{\cal{H}}(\mathbf{r}) = \frac{\hbar^2}{2m}\tau_0
+ {\cal H}_0(\mathbf{r})
+ {\cal H}_1(\mathbf{r})
+ {\cal H}_C(\mathbf{r})~.
\end{equation}
The isoscalar and isovector terms carry an isospin index ($I = 0, 1$) for the energy densities, respectively. The isoscalar (${\cal H}_0(\mathbf{r})$) energy density depends on the isoscalar particle density, $\rho_0 = \rho_n + \rho_p$, whereas the isovector (${\cal H}_1(\mathbf{r})$) energy density depends on the isovector particle
density, $\rho_1 = \rho_n - \rho_p$. These definitions, of course, prescribe analogous expressions for other densities and
currents.
The local gauge and Galilean invariant form is given by~\cite{dobaczewski1995}
\begin{equation}
\label{eq:efunctional}
\begin{split}
{\cal H}_\mathrm{I}(\mathbf{r})
& = C_\mathrm{I}^{\rho}            \rho_\mathrm{I}^2
+  C_\mathrm{I}^{   s}            \mathbf{s}_\mathrm{I}^2
+  C_\mathrm{I}^{\Delta\rho}      \rho_\mathrm{I}\Delta\rho_\mathrm{I} \\
&+  C_\mathrm{I}^{\Delta s}        \mathbf{s}_\mathrm{I}\cdot\Delta\mathbf{s}_\mathrm{I}
+  C_\mathrm{I}^{\tau}      (\rho_\mathrm{I}\tau_\mathrm{I}-\mathbf{j}_\mathrm{I}^2) \\
&+  C_\mathrm{I}^{   T}      \Big(\mathbf{s}_\mathrm{I}\cdot
\mathbf{T}_\mathrm{I} - \tensor{J}_\mathrm{I}^2\Big)
+ C_\mathrm{I}^{\nabla J}  \Big(\rho_\mathrm{I}\mathbf{\nabla}\cdot\mathbf{J}_\mathrm{I}
+ \mathbf{s}_\mathrm{I}\cdot
(\mathbf{\nabla}\times\mathbf{j}_\mathrm{I})\Big)~.
\end{split}
\end{equation}
The density dependence of the coupling constants has been restricted to the $C_\mathrm{I}^{\rho}$ and $C_\mathrm{I}^s$ terms only
which stems from the most common choice of Skyrme EDF.
These density dependent coefficients contribute to the coupling of isoscalar and isovector fields
in the Hartree-Fock Hamiltonian~\cite{dobaczewski1995}.

The decomposition of the Skyrme EDF into isoscalar and isovector components makes it feasible to study isospin dependence of nuclear properties microscopically, both for nuclear reactions~\cite{vophuoc2016,godbey2017} as well as for nuclear structure~\cite{dobaczewski1995}.
This is possible for any approach that employs the Skyrme EDF to compute ion-ion interaction potentials. Here,
we implement the decomposed Skyrme EDF in the density-constrained DC-TDHF method~\cite{umar2006b,godbey2017} to study
isospin effects in fusion barriers.
The DC-TDHF approach permits the study of sub-barrier fusion through the direct calculation of nucleus-nucleus potentials, $V(R)$,
from TDHF dynamics. The DC-TDHF method has been used in the study of fusion for a wide range of nuclear reactions~\cite{umar2006a,umar2014a,simenel2013a,umar2012a,keser2012,oberacker2010,jiang2015a}.
The basic idea of the DC-TDHF method is the following:
At certain time steps, $t$ (or, internucleon distances $R(t)$), a minimization of the static energy
is performed while proton and neutron densities are constrained to be the instantaneous densities yielded from the TDHF equations. That is,
\begin{equation}
\begin{split}
    E_{DC}(R) &= \bigg\{E[\rho_n,\rho_p] + \int d^3r \lambda_n(\mathbf{r})[\rho_n(\mathbf{r}) - \rho_n^{tdhf}(\mathbf{r}, t)] \\
              &+ \int d^3r\lambda_p(\mathbf{r})[\rho_p(\mathbf{r}) - \rho_p^{tdhf}(\mathbf{r}, t)]\bigg\}\bigg|_{min_{\rho}}
\end{split}
\end{equation}
where $\lambda_n(\mathbf{r})$ and $\lambda_p(\mathbf{r})$ are Lagrange multipliers. This minimized energy is referred to as the so-called `density constrained energy',  $E_{\mathrm{DC}}(R)$. In essence, all excitation energy has been removed from the system through this procedure. To obtain the underlying ion-ion interaction potential, $V(R)$, the constant binding energies (obtained from a static Hartree-Fock approach) of the two individual nuclei ($E_{\mathrm{A_{1}}}$ and $E_{\mathrm{A_{2}}}$) are then subtracted
\begin{equation}
V(R)=E_{\mathrm{DC}}(R)-E_{\mathrm{A_{1}}}-E_{\mathrm{A_{2}}}\ .
\label{eq:vr}
\end{equation}
Ion-ion interaction barriers calculated from the DC-TDHF approach self-consistently contain all of the dynamical changes in the nuclear density throughout the TDHF reaction.
Utilizing the decomposition of the Skyrme EDF [Eq.~(\ref{eq:efunctional})], we can re-write this potential as
\begin{equation}
V(R) = \sum_{\mathrm{I}=0,1} v_\mathrm{I}(R) + V_C(R)~,
\end{equation}
where $v_\mathrm{I}(R)$ denotes the potential computed by using the isoscalar and isovector parts of
the Skyrme EDF given in Eqs.~(\ref{eq:edensity}) and (\ref{eq:vr}).
The Coulomb potential is solved from the typical three-dimensional Poisson equation (where the Slater approximation is used for the Coulomb exchange term) via Fast-Fourier Transform techniques.
\begin{figure}[!htb]
\includegraphics*[width=8.6cm]{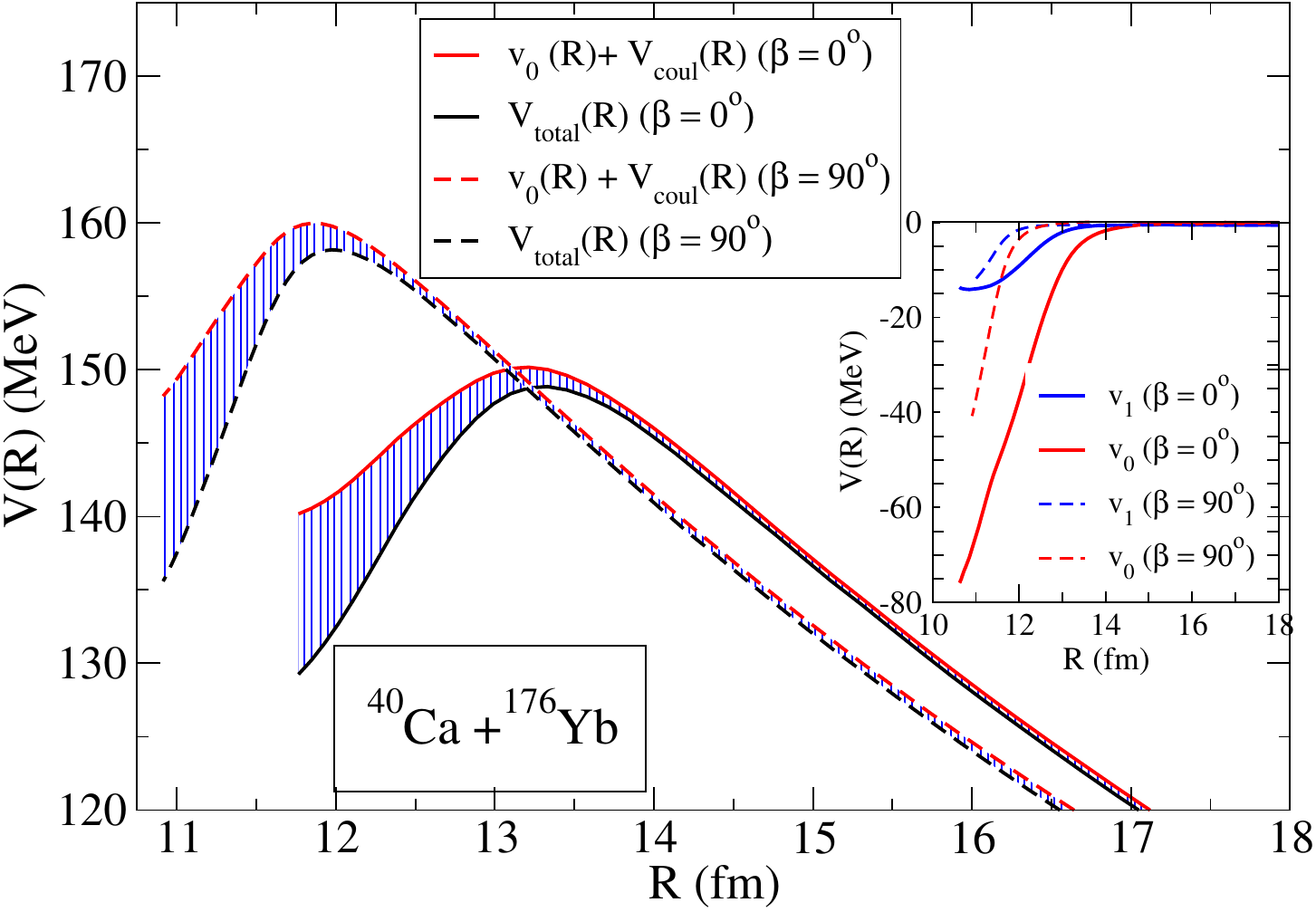}
\caption{For the $^{40}$Ca+$^{176}$Yb system:
Total and isoscalar DC-TDHF potentials for two orientations of the prolate-deformed $^{176}$Yb (dashed lines denote a Euler angle rotation of $\beta = 90^\circ$). The shaded region in blue depicts a significant \textit{reduction} as an effect of the isovector contribution to the energy density. The insert shows the isoscalar and isovector contributions to the interaction barrier without the Coulomb potential. The TDHF collision energy was $E_\mathrm{c.m.}=166.45$~MeV.}
\label{fig:Ca40Yb176}
\end{figure}

We have implemented the DC-TDHF approach to study fusion barriers for a number of systems involving spherical isotopes of Calcium without the use of the pairing interaction (in particular, Calcium-40, -44, -48, and -54) on prolate-deformed Ytterbium-176, which permits us to also inspect the orientation dependence of isospin flow.
All calculations were done on a three-dimensional
Cartesian lattice with no symmetry assumptions~\cite{umar2006c}, and the
Skyrme SLy4d EDF~\cite{kim1997} was used.
The Cartesian box size utilized for all calculations was chosen to be $60\times 32\times 32$~fm$^3$, with a mesh spacing of
$1.0$~fm in all directions. Employing an advanced numerical discretization technique known as the basis-spline collocation method~\cite{umar1991a}, these values provide very accurate numerical results.
\begin{figure}[!htb]
\includegraphics*[width=8.6cm]{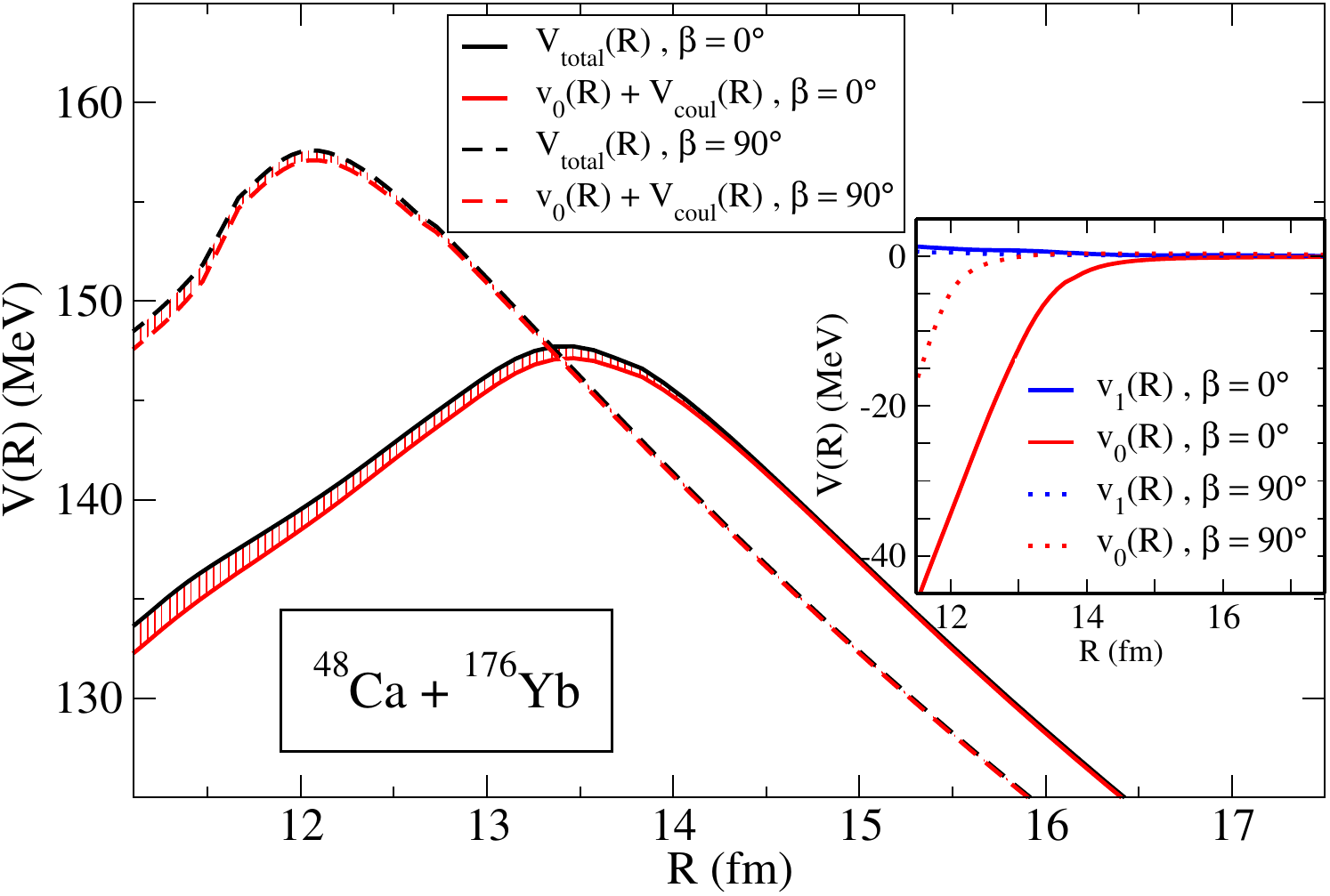}
\caption{For the $^{48}$Ca+$^{176}$Yb system:
Total and isoscalar DC-TDHF potentials for two orientations of the prolate-deformed $^{176}$Yb (dashed lines denote a Euler angle rotation of $\beta = 90^\circ$). The shaded region in red depicts a small \textit{enhancement} as an effect of the isovector contribution to the energy density. The insert shows the isoscalar and isovector contributions to the interaction barrier without the Coulomb potential. The TDHF collision energy was $E_\mathrm{c.m.}=161$~MeV.}
\label{fig:Ca48Yb176}
\end{figure}

For each system under consideration, separate DC-TDHF calculations have been performed for two orientations of the prolate-deformed $^{176}$Yb nucleus: Euler angle rotations corresponding to $\beta = 0^\circ$ and $\beta = 90^\circ$ (solid and dashed lines, respectively). The center-of-mass energy was chosen to be 1.05 times the corresponding Bass barrier for each system.
We begin with fusion of $^{40}$Ca + $^{176}$Yb, colliding at E$_{c.m.}$ = 166.45 MeV plotted in Fig.~\ref{fig:Ca40Yb176}.
The black curves denote the total DC-TDHF potential while the red curves are the combination of isoscalar and Coulomb
potentials. The difference between these curves show the net isovector contribution to the ion-ion interaction potential
(shaded regions).
For the symmetric, doubly magic $^{40}$Ca nucleus colliding with either orientation of $^{176}$Yb, there is a substantial reduction of the barrier (area shaded in blue) as a result of the added isovector potentials.
This we refer to as the \textit{isovector reduction}, meaning that the isovector contribution is making the overall
potential thinner and in the inner barrier region with a slightly lower barrier height.
The inset graph shows the isoscalar/isovector potential contributions by themselves, without the Coulomb energy.
\begin{figure}[!htb]
\includegraphics*[width=8.6cm]{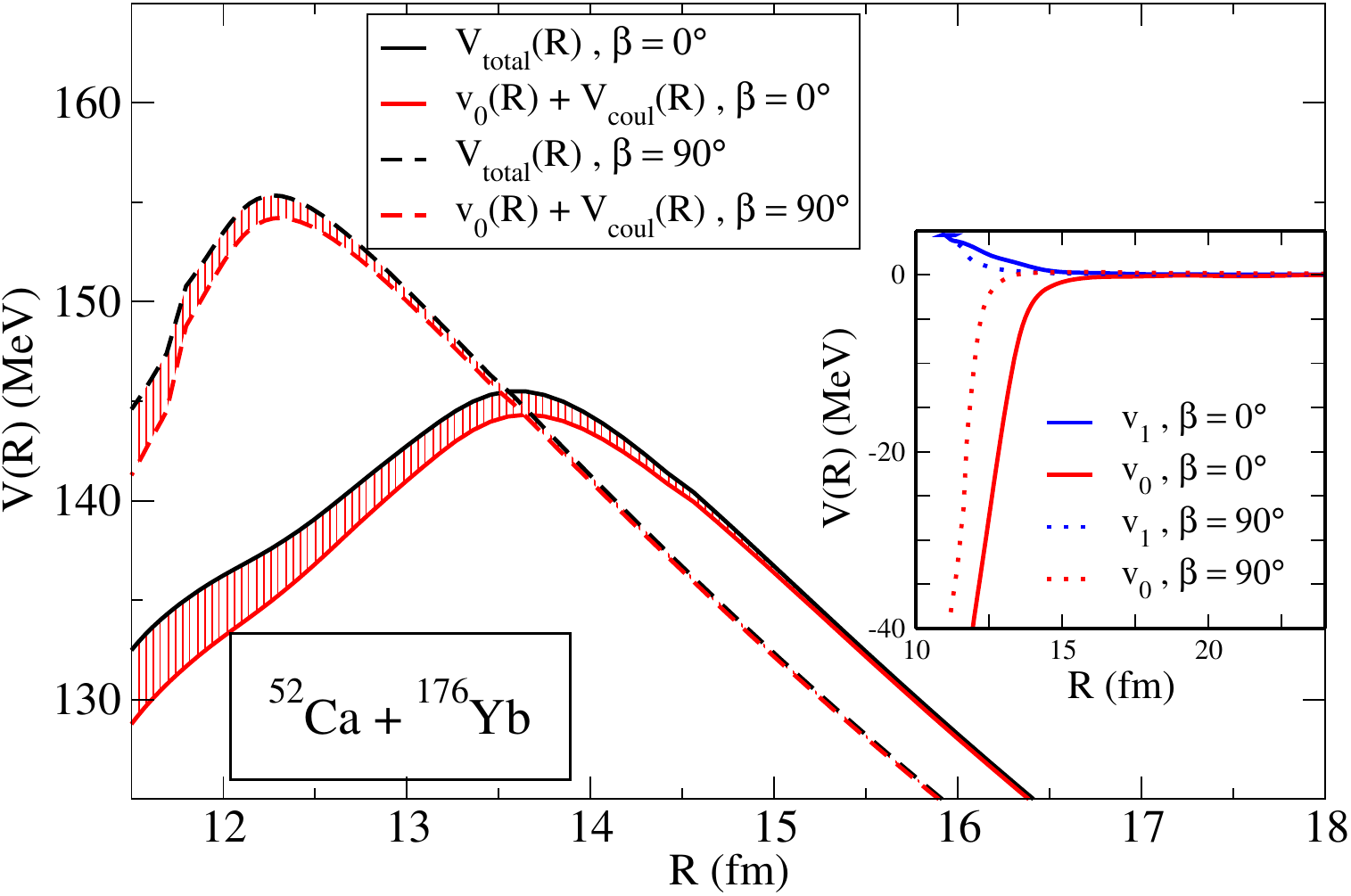}
\caption{For the $^{52}$Ca+$^{176}$Yb system:
Total and isoscalar DC-TDHF potentials for two orientations of the prolate-deformed $^{176}$Yb (dashed lines denote a Euler angle rotation of $\beta = 90^\circ$). The shaded region in red depicts a significant \textit{enhancement} as an effect of the isovector contribution to the energy density. The insert shows the isoscalar and isovector contributions to the interaction barrier without the Coulomb potential. The TDHF collision energy was $E_\mathrm{c.m.}=159.9$~MeV.}
\label{fig:Ca52Yb176}
\end{figure}
\begin{figure}[!htb]
\includegraphics*[width=8.6cm]{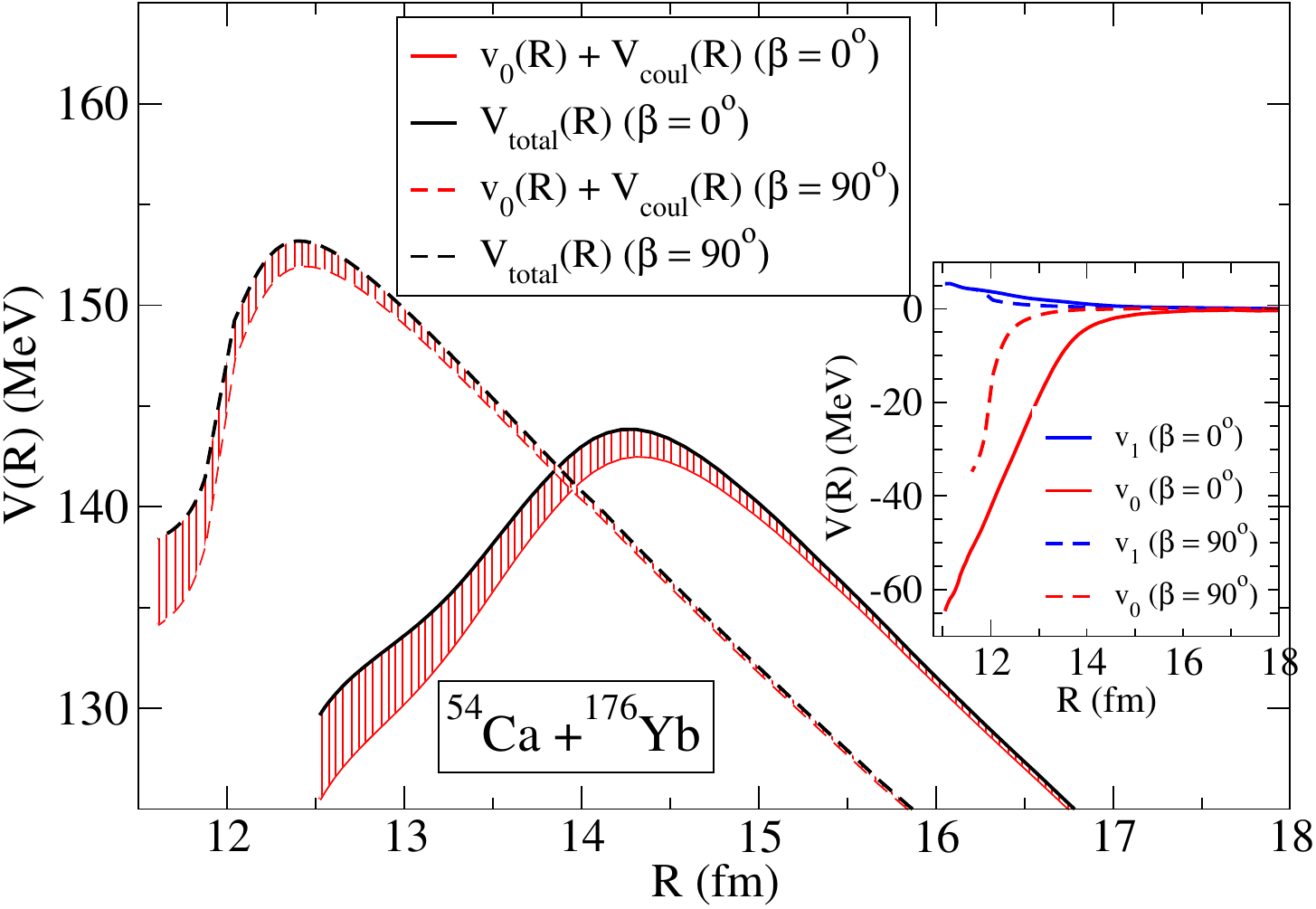}
\caption{For the $^{54}$Ca+$^{176}$Yb system:
Total and isoscalar DC-TDHF potentials for two orientations of the prolate-deformed $^{176}$Yb (dashed lines denote a Euler angle rotation of $\beta = 90^\circ$). The shaded region in red depicts an even more significant \textit{enhancement} as an effect of the isovector contribution to the energy density. The insert shows the isoscalar and isovector contributions to the interaction barrier without the Coulomb potential. The TDHF collision energy was $E_\mathrm{c.m.}=158.98$~MeV.}
\label{fig:Ca54Yb176}
\end{figure}

Next, we examine the $^{48}$Ca+$^{176}$Yb system at E$_{c.m.}$ = 161 MeV. In Fig.~\ref{fig:Ca48Yb176}, with the addition of 8 neutrons to the system, we start to see the role of the isovector contribution to the energy density change.  Rather than the reduction observed in with $^{40}$Ca, there is now a small \textit{ isovector enhancement} of the potential barrier (areas shaded in red). This difference in potential barriers for the $^{40}$Ca and $^{48}$Ca is analogous to the experimental observation of a sub-barrier fusion enhancement in the system $^{132}$Sn+$^{40}$Ca as compared
to more neutron-rich system $^{132}$Sn+$^{48}$Ca~\cite{kolata2012}.
It was shown in an earlier publication~\cite{oberacker2013} that for most systems isovector dynamics results in the thinning of
the barrier thus enhancing the sub-barrier fusion cross-sections.
The isovector reduction effect vanishes for symmetric systems as well
as the $^{48}$Ca+$^{132}$Sn system for which neutron pickup $Q$-values are all negative.
This enhancement effect becomes more pronounced as further neutrons are introduced to the calcium nuclei.  For $^{52}$Ca+$^{176}$Yb at E$_{c.m.}$ = 159.9 MeV (Fig.~\ref{fig:Ca52Yb176}) and then $^{54}$Ca+$^{176}$Yb at E$_{c.m.}$ = 158.98 MeV (Fig.~\ref{fig:Ca54Yb176}), the potentials calculated from solely the isoscalar and Coulomb terms are now both lower in peak energy, and smaller in width than those calculated with the total density functional.
\begin{figure}[!htb]
\includegraphics*[width=8.6cm]{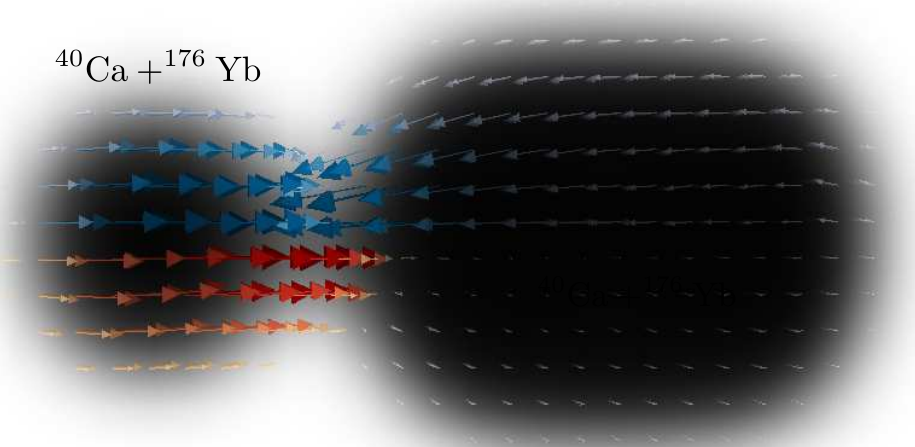}
\caption{For the $^{40}$Ca+$^{176}$Yb system: Single-particle currents for neutrons
(upper half slice shown in blue) and for protons (lower half slice shown in red).
For this system we observe that the net neutron flow is from $^{176}$Yb to
$^{40}$Ca, while the proton flow is in the opposite direction.
Also shown is the shaded outline of the
position of the two nuclei (in this case for the $\beta=0^{\circ}$ orientation of $^{176}$Yb).}
\label{fig:40Ca176Yb-current}
\end{figure}
\begin{figure}[!htb]
\includegraphics*[width=8.6cm]{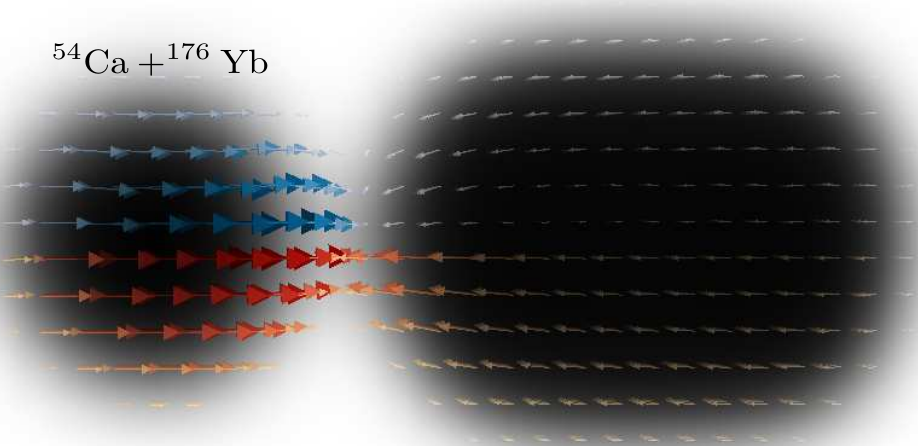}
\caption{For the $^{54}$Ca+$^{176}$Yb system: Single-particle currents for neutrons
(upper half slice shown in blue) and for protons (lower half slice shown in red).
For this system we observe that the net neutron and proton flow is from $^{54}$Ca
to $^{176}$Yb.Also shown is the shaded outline of the
position of the two nuclei (in this case for the $\beta=0^{\circ}$ orientation of $^{176}$Yb).}
\label{fig:54Ca176Yb-current}
\end{figure}

In all the reactions studied here, we also note that the effect of isovector contribution
is more enhanced for the tip orientation ($\beta=0$) of the target nucleus.
This is likely due to the fact that the contact with the tip orientation happens
earlier (larger $R$) compared to the side orientation. Since the side orientation
normally would have a larger area of contact with the projectile this suggests a
competition between time spent between the two nuclei prior to fusion and the
size of the overlap region. Thus, nucleon transfer should also depend on the
orientation for deformed nuclei, which is normally not taken into account in
non-microscopic approaches.
It is possible to provide a further insight to these results by examining the transfer of neutrons
and protons during the contact phase of the collision process since the isovector contribution is
intimately related to transfer properties. For this end we have plotted the single-particle currents during
the TDHF evolution. In Fig.~\ref{fig:40Ca176Yb-current} we plot these currents at the initial
contact phase for the $^{40}$Ca+$^{176}$Yb system, together with the shaded outline of the
position of the two nuclei (in this case for the $\beta=0^{\circ}$ orientation of $^{176}$Yb).
The upper half-plane shows the direction of neutron flow (blue arrows) while the lower half-plane
shows the proton currents (red arrows). We observe that in this case neutrons are flowing from the
$^{176}$Yb towards $^{40}$Ca, while the proton flow is in the reverse direction from $^{40}$Ca towards
$^{176}$Yb. This mode of transfer leads to the isovector reduction of the potential barrier.
 In Fig.~\ref{fig:54Ca176Yb-current} we plot the same quantities for the $^{54}$Ca+$^{176}$Yb system.
 In this case we observe that both neutrons and protons are flowing from $^{54}$Ca to $^{176}$Yb.
 The case for $^{52}$Ca is similar to the $^{54}$Ca and for the $^{48}$Ca the net transfer is
 negligibly small, which explains why the there is very little isovector contribution to the
 fusion barrier.

 In summary, we have performed DC-TDHF calculations with the decomposed EDF into isoscalar and
 isovector parts with the purpose of identifying the isovector contribution to the overall
 fusion potential barrier. The isovector contribution is an indicator of the influence of particle
 transfer during the early stages of nuclear contact prior to fusion. We observed that for the
 $^{40}$Ca+$^{176}$Yb system the neutron transfer is from the target to projectile, which leads
 to the reduction of the potential barrier, whereas for the neutron rich systems the transfer
 reverses direction and leads to the enhancement of the potential barrier. These changes effect
 both the height and the width of the barrier. We also observe that transfer does depend on the
 orientation of the deformed target.

\begin{acknowledgments}
This work has been supported by the U.S. Department of Energy under award numbers DE-SC0013847 (Vanderbilt University).  In addition, co-author Richard Gumbel acknowledges support from a Fisk-Vanderbilt Master's-to-PhD Bridge Program fellowship.

\end{acknowledgments}

\bibliography{VU_bibtex_master}

\end{document}